\documentclass[utf8]{frontiersSCNS} 

\usepackage{url,hyperref,lineno,microtype,subcaption}
\usepackage[onehalfspacing]{setspace}
\newcommand{\diamonds}{\textsc{Diamonds}}
\newcommand{\kepler}{\textit{Kepler}}

\def\keyFont{\fontsize{8}{11}\helveticabold }
\def\firstAuthorLast{Corsaro E.} 
\def\Authors{Enrico Corsaro\,$^{*}$}
\def\Address{$^{}$INAF -- Osservatorio Astrofisico di Catania, via S. Sofia 78, 95123 Catania, Italy}
\def\corrAuthor{Dr. Enrico Corsaro}

\def\corrEmail{enrico.corsaro@inaf.it}

\begin{document}
\onecolumn
\firstpage{1}

\title[Oscillation frequency extraction with Bayesian multi-modality]{Fast and automated oscillation frequency extraction using Bayesian multi-modality} 

\author[\firstAuthorLast ]{\Authors} 
\address{} 
\correspondance{} 

\extraAuth{}

\maketitle

\begin{abstract}
Since the advent of CoRoT, and NASA \textit{Kepler} and K2, the number of low- and intermediate-mass stars classified as pulsators has increased very rapidly with time, now accounting for several $10^4$ targets. With the recent launch of NASA TESS space mission, we have confirmed our entrance to the era of all-sky observations of oscillating stars. TESS is currently releasing good quality datasets that already allow for the characterization and identification of individual oscillation modes even from single 27-days shots on some stars. When ESA PLATO will become operative by the next decade, we will face the observation of several more hundred thousands stars where identifying individual oscillation modes will be possible. However, estimating the individual frequency, amplitude, and lifetime of the oscillation modes is not an easy task. This is because solar-like oscillations and especially their evolved version, the red giant branch (RGB) oscillations, can vary significantly from one star to another depending on its specific stage of the evolution, mass, effective temperature, metallicity, as well as on its level of rotation and magnetism. In this perspective I will present a novel, fast, and powerful way to derive individual oscillation mode frequencies by building on previous results obtained with \diamonds. I will show that the oscillation frequencies obtained with this new approach can reach precisions of about 0.1\,\% and accuracies of about 0.01\,\% when compared to published literature values for the RGB star KIC~12008916.

\tiny
 \keyFont{ \section{Keywords:} asteroseismology, oscillation mode analysis, stars: oscillations, stars: late-type, methods: statistical, methods: numerical, methods: data analysis, Bayesian}
\end{abstract}

\section{Introduction}
Despite both CoRoT \citep{Baglin06} and NASA \kepler\,and K2 missions \citep{Borucki10,Koch10,Howell14K2} have ended, we keep gathering new datasets of oscillating stars from recently launched NASA TESS \citep{Ricker14TESS}, which is offering the opportunity to extract individual oscillation mode frequencies even with 27-days long observations (see e.g. Huber et al. submitted). The rapidly growing amount of stars with visible stellar oscillation modes, expected to encounter a significant step increase with the advent of the future ESA PLATO space mission \citep{Rauer14PLATO}, poses the problem of facing an adequate type of analysis to fully characterize stellar oscillations on a large sample basis. The full characterization of the oscillation modes is a high priority task because it delivers the largest amount of most valuable information that we can extrapolate from each star, essential to understand stellar structure and evolution in detail \citep{Aerts10}.

First works focusing on extracting and exploiting oscillation mode frequencies, amplitudes, and linewidths were already carried out \citep[see e.g.][]{Corsaro15cat,Corsaro15glitch,Benomar15,Davies16,Lund17LEGACY,Corsaro17Nature,Handberg17,Garcia18pb,Vrard18pb}, showing that an effort to tackle the problem is currently ongoing. Nevertheless the analysis proves to be challenging because of the large number of oscillation modes involved, although it is still possible to automatize it in some cases.

My view on the problem is that we require a simple approach that can be at the same time powerful, flexible to be adapted to different conditions, and fast in performing a peak bagging analysis for each star. This approach should also be accessible by the global asteroseismic community. For this purpose, I present a novel methodology, based on the public code \diamonds\,\,\citep[][hereafter C14]{Corsaro14}, to extract individual oscillation mode frequencies that is at the same time fast, easy to use, and accurate. For demonstrating its working principle and performance, I apply it to the RGB star KIC~12008916, for which a detailed peak bagging analysis is available \citep[][hereafter C15]{Corsaro15cat}.

\section{Fast oscillation frequency extraction with \diamonds}
\label{sec:fast}
An efficient and robust approach to the fitting of individual oscillation modes exploits the Bayesian inference based on the nested sampling Monte Carlo algorithm \citep[][C14]{Skilling04,Sivia06}. This is because the nested sampling has the advantages of: 1) requiring a factor 100 less sampling points than standard Markov chain Monte Carlo methods to accurately perform a statistical inference; 2) computing the Bayesian evidence of the model, a key element to test models in the light of both their complexity and fit quality; 3) efficiently sampling likelihood distributions that exhibit phase changes, i.e. multiple local maxima.

With the development of \diamonds\footnote{https://github.com/EnricoCorsaro/DIAMONDS}, a multi-platform publicly available Bayesian inference tool in {\ttfamily C++11} \citep[see also C14; C15;][for general guidelines]{Corsaro18tutorial}, the characterization of the multitude of oscillation modes typical of the low- and intermediate-mass stars has been made accessible through the nested sampling and the flexible code implementation. In C14 two completely different approaches to peak bagging for a main sequence star were presented, which I summarize below. 

\begin{enumerate}
\item The first approach, uni-modal and high-dimensional (hereafter Approach 1), requires priors set up for each of the individual mode properties of frequency centroid, amplitude and linewidth. The resulting sampling of the parameter space will exhibit a single global maximum of the likelihood distribution. Approach 1 was also used extensively to perform peak bagging analysis in about 90 red giant stars \citep[C15;][]{Corsaro17Nature}. This procedure yields the most precise and accurate set of asteroseismic measurements, once appropriate prior distributions are obtained. However, the fit is necessarily high-dimensional (with three free parameters for each oscillation mode being included in the fit), and although \diamonds\,\,can still afford a peak bagging fit up to about $k\sim40$ dimensions without particular problems \citep[see the new calibration for peak bagging by][]{Corsaro18corr}, the computational efficiency as well as the computational time required to carry out an individual fit increases quite rapidly with increasing number of oscillation modes being fitted, with a time dependency that follows a $k^{2.4}$ power law\footnote{See the \diamonds\,\,2018 User Guide Manual available in the same GitHub repository of \diamonds\,\,for more details.}. While this aspect can be overcome by dividing the power spectrum of the star into multiple chunks, which are considered and analyzed independently from one another, one still has to face the problem of setting up priors for each oscillation peak that should be fitted, which in practice represents the main limitation. 
\item The second approach, originally proposed by C14, is multi-modal and low-dimensional (hereafter Approach 2). The initial recipe accounted for a model with three Lorentzian profiles trying to reproduce the $\ell = 2,0,1$ triplet typical of main sequence solar-like pulsators. This was done through the fitting of the small frequency spacings between adjacent $\ell = 2,0$ modes and adjacent $\ell = 0, 1$ modes and of a frequency centroid $\nu_1$ of the dipolar mode used to locate the triplet of peaks in the spectrum. With this peak bagging model, the resulting sampling exhibits a series of clustered regions in the parameter space, each one corresponding to a local maximum (island) of the likelihood \citep[see Figure 13 of ][for a visualization]{Corsaro14}. The approach proved to be able to recover the oscillation mode properties of 27 consecutive peaks of a main sequence star using only nine free parameters, with an average accuracy on the oscillation frequencies of about 0.2\,\% as compared to estimates from Approach 1 for the same star. However, setting up \diamonds\,\,for this configuration appeared not to be straightforward, and even the computational time required to perform one run (about 1 hour in a 2.6 GHz CPU core for a \kepler\,\,short cadence power spectrum accounting for 27 oscillation peaks) did not yield any significant gain with respect to performing Approach 1 to the same frequency region. 
\end{enumerate}

In the following sections, I describe how to revise the original Approach 2 presented by C14 to improve significantly its speed, simplicity in use, and also its reliability in producing accurate and precise estimates of individual oscillation frequencies.

\subsection{Islands peak bagging model}
The choice of the model to fit the so-called power spectral density (PSD), namely the power spectrum normalized by the frequency resolution of the dataset, requires that the output sampling from \diamonds\,\,has to resemble that of a multi-modal likelihood distribution. In order to create a model that is at the same time as simple as possible so that it can be easily configured, and also effective in recognizing and fitting the actual oscillation peaks that are present in the stellar power spectra, I consider a peak bagging model consisting of a single Lorentzian profile
\begin{equation}
P_\mathrm{isla} \left(\nu ; \Gamma \right) = \frac{A^2/(\pi \Gamma)}{1 + 4\left( \frac{\nu - \nu_0}{\Gamma}\right)^2}
\end{equation}
which I refer to as the \emph{islands} model, where $A$ is the amplitude of the oscillation peak, $\Gamma$ its FWHM, and $\nu_0$ its frequency centroid. The model is superimposed on a fixed background level that can be fitted in a previous step \citep[see e.g.][for a formulation]{Corsaro17metallicity}. It is important to note that in the islands model only $A$ and $\nu_0$ are considered as free parameters since $\Gamma$ is fixed to a value that is related to the actual $\nu_\mathrm{max}$ of the star. By fixing the FWHM of the peak we can stabilize the fit and obtain a better resolving power of the actual oscillation peak structures that are present in the dataset. The choice of $\Gamma$ proves not to be particularly critical, but the general idea behind is that it should be set to a value equal to or below the minimum linewidth of an oscillation peak among those of a given star. I refer to Section~\ref{sec:discussion} for more discussion about the choice of $\Gamma$ for the application presented in this work. With this islands model, we therefore account for only two free parameters, clearly expecting a considerable gain in speed when performing the fit. In addition, because of the adoption of just one Lorentzian profile and the only need for a rough assumption about the linewidth and amplitude of the observed peak structures, the sampling from \diamonds\,\,proves to be significantly more stable with respect to the original Approach 2 presented by C14, thus producing more reliable results.

\subsection{Multi-modal sampling at high threshold}
Once the islands model is constructed, one can perform the actual fit to the PSD by providing as uniform prior distribution for the frequency centroid $\nu_0$ an input range that spans the actual region of the PSD that we intend to analyze, and a uniform prior range for the amplitude that resembles the average level of the amplitude of the peaks in the spectrum. This is the way we practically make the islands model a peak bagging model producing a multi-modal likelihood distribution. In the test performed, an input range for $\nu_0$ comparable to the large frequency separation $\Delta\nu$ offers an optimal choice in terms of resulting frequency resolution of the sampling for detecting all the oscillation peaks reported by C15.

Before performing the fit, it is essential to provide some considerations on the parameters that configure the nested sampling algorithm of \diamonds\,\,\citep[see C14; C15; ][for more details]{Corsaro18tutorial, Corsaro18corr}. In \cite{Corsaro18corr} in particular, it was proved that using a number of $N_\mathrm{live} = 500$ live points is adequate for Approach 1 to converge to an accurate solution even if a large number of dimensions is involved (up to $k\sim60$). This relatively low number of live points can sustain the fit because of the uni-modality nature of the problem. In a multi-modal application instead, as indicated by C14, we require a high resolution of live points in the parameter space to be able to locate the individual local maxima of the likelihood distribution. This is because we know a priori that the likelihood will contain several, if not many, different local maxima. In this regard, C14 adopted a value of $N_\mathrm{live} = 2000$ and performed the fit until the termination condition of $\delta_\mathrm{final} = 0.01$ was reached \citep[see also][fore more discussion]{Keeton11}. While the choice of the number of live points seems adequate also for the new application, that of adopting a low threshold $\delta_\mathrm{final}$ constitutes a severe problem. The result is that not only the fit takes a long time to be performed in its entirety (many nested iterations are required), but that it is also very likely that the computation will fail before reaching the end because the sampling has to take place from all or most of the local maxima at each attempt to draw a new sampling point. Clearly, the more we approach to the threshold, the more difficult it will be to find a new point that satisfies the new higher likelihood constraint. 

\begin{figure}[h!]
\begin{center}
\includegraphics[width=17.7cm]{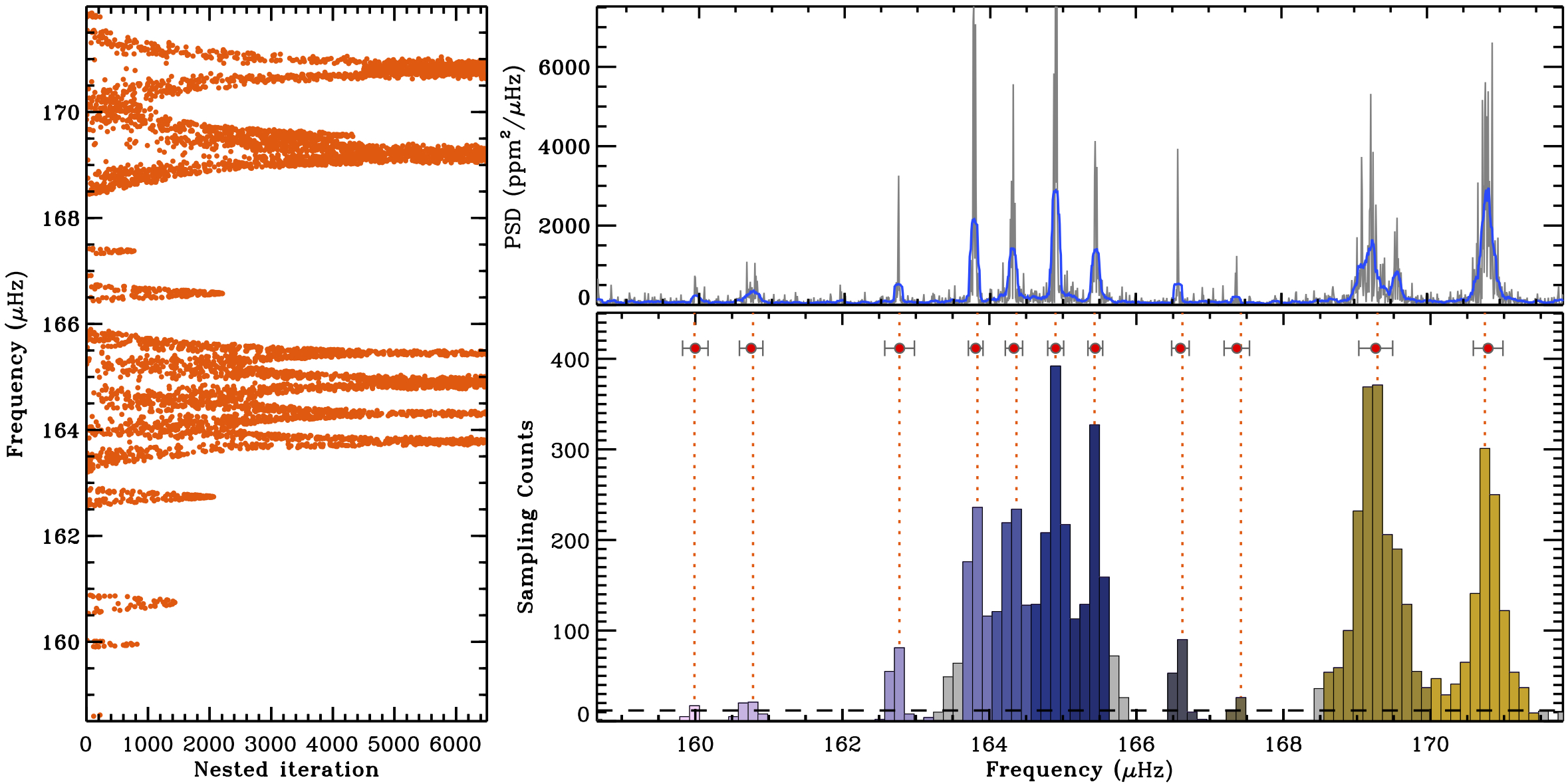}
\end{center}
\caption{\textbf{Left panel:} Sampling evolution from \diamonds\,\,of the frequency centroid $\nu_0$ (orange dots) as a function of the nested iteration, for the red giant KIC~12008916. Each nested iteration corresponds to one sampling point. The sampling presented is built as follows: 1) \diamonds\,\,performed a total of $N_{\rm nest} = 6000$ iterations (hence 6000 sampling points) using 2000 live points; 2) the run is completed by adding up the set of 2000 live points that is left at the end of the nested sampling, thus resulting in a total of 8000 effective nested iterations (or equivalently sampling points); 3) the sampling from the first 1500 nested iterations was removed to improve the detection of the peak structures in the PSD. This results in a final set of 6500 nested iterations, hence of sampling points used for the analysis. \textbf{Right panels:} (top) The PSD of the corresponding chunk analyzed (in gray) with a 10 frequency-bin smoothing overlaid (blue curve) to highlight the oscillation peak structures. (bottom). The counts histogram from the sampling, with local maxima corresponding to the extracted oscillation frequencies indicated by vertical dotted red lines, as identified by a hill climbing algorithm. The different regions of the histogram, denoted with varying colors, represent the ranges of frequency automatically selected to compute the final estimates of the oscillation frequencies and their corresponding uncertainties, according to the method (b) presented in Figure~\ref{fig:2}. The estimated frequencies and corresponding 1-$\sigma$ uncertainties from method (b) are indicated by the red circles and error bars, respectively. The horizontal dashed line represents the threshold level for the selection of the peaks, set to 3\,\% of the maximum counts in the histogram.}
\label{fig:1}
\end{figure}

For the reasons just presented, I have implemented another termination condition in \diamonds\,\,that can take into account a fixed number of nested iterations. This allows to decide when to stop the computation depending on how many sampling points have been drawn. In this way, one is independent of the actual evidence estimated from the sampling, which can change significantly from one fit to another. Given that each nested iteration progressively corresponds to an increasing likelihood value, the peaks in the PSD that have a low height compared to the level of the background, quickly disappear after popping up in the sampling, while the peaks that have a larger height in the PSD are sampled for a longer time because they yield a larger likelihood value to the fit. This can be  visualized in the left panel of Figure~\ref{fig:1}, for the application discussed in Section~\ref{sec:application}, where the sampling from \diamonds\,\,terminates for five different peaks after about 2000 iterations since being sampled\footnote{The nested iteration number reported in Figure~1 does not correspond to the real nested iteration number used by \diamonds\,\,during its sampling. The reader can refer to the caption of Figure~\ref{fig:1} for more details on how this final sampling is obtained.}, while continuing until the end for the other six peaks. This effect can be summarized as follows: 1) if we stop too late in the nested sampling process we tend to oversample those peaks that are bigger, thus killing out all the structures that have a small height in the PSD; 2) if we stop too early, we risk to visualize too many structures that belong to the noise realization of the dataset, thus crowding with noise peaks the actual set of oscillation peaks that we can extrapolate from the sampling. Adopting $N_\mathrm{live} = 2000$ and a high threshold stopping point, set to $N_\mathrm{nest} = 6000$, produces a very stable and reliable sampling of the structures observed in the PSD for the application presented in this work. All the other configuring parameters of \diamonds\,\,remain unchanged and can be set according to \cite{Corsaro18corr}.

\subsection{Counts histogram for the extraction of individual oscillation frequencies}
By using the sampling cloud obtained from \diamonds\,\,for the free parameter $\nu_0$, we can proceed with the construction of a corresponding counts histogram. I adopt a resolution per bin in $\nu_0$ given by 1/100 of the length of the parameter range, where each bin in frequency will contain the number of sampling points falling in the given bin. This proves to be adequate to detect at least 11 different peaks within a frequency range of length $\Delta\nu$ and to get rid of spurious structures arising from the noise. As a result, when the sampling matches an actual oscillation peak in the PSD, a corresponding peak pops up in the counts histogram. The subsequent step is to consider a detection threshold in counts in order to pick up only those peaks of the histogram that exceed the threshold. A value for the threshold of 3\% of the maximum number of counts found in the histogram appear to provide an optimal condition for the test presented in this work in order to detect all the oscillation peaks reported by C15. Once the threshold is defined, one can rely on a relatively simple hill climbing algorithm to progressively gather the local maxima present in the histogram by starting from the left edge of the frequency range. The result for this application is shown in the right panel of Figure~\ref{fig:1}. The adoption of a counts histogram coupled with a hill climbing algorithm is an important step because it provides a stable visualization and identification of the peak structures sampled from the PSD. This in turn makes the process of extracting the oscillation frequencies fast, automated, and reliable, in contrast to the original Approach 2 presented by C14 where the actual peak structures had to be manually identified from the sampling cloud obtained by \diamonds.

\subsection{Comparison with Approach 1 for RGB oscillations}
\label{sec:application}
\begin{figure}[h!]
\begin{center}
\includegraphics[width=17.7cm]{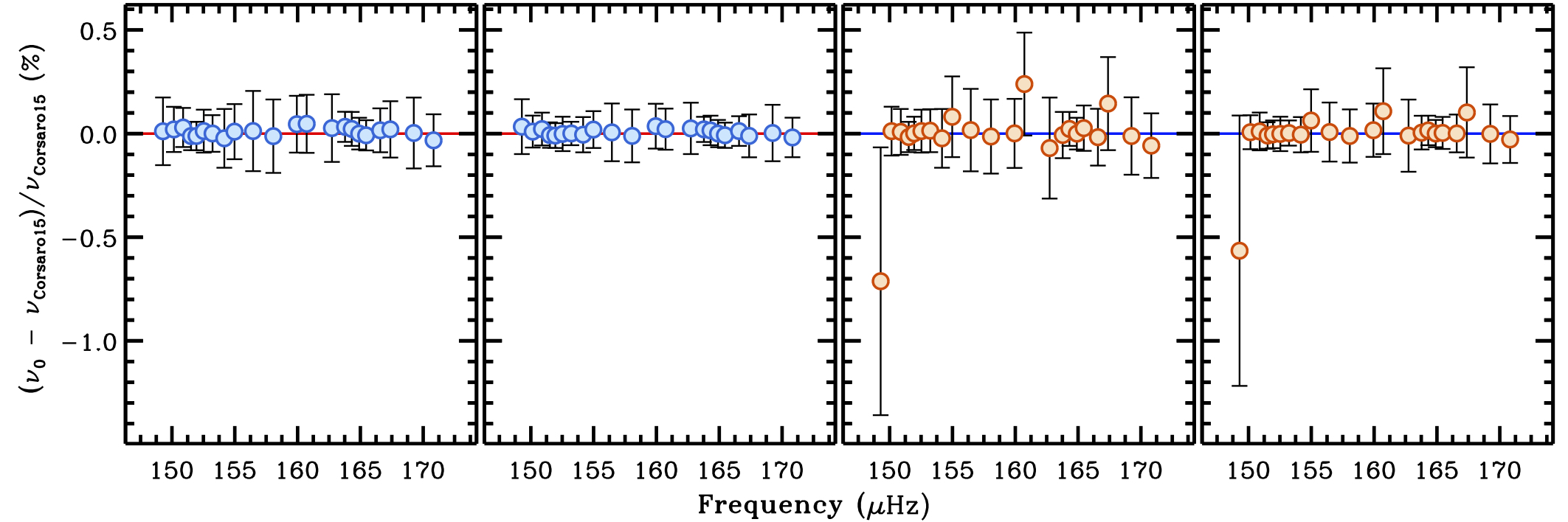}
\end{center}
\caption{Comparison between the oscillation frequencies extracted from \diamonds\,\,using the multi-modal approach ($\nu_0$) and those published by C15 ($\nu_\mathrm{Corsaro15}$). The values are reported in percentage of deviation with respect to the literature values. The uncertainty error bars obtained with the new approach are also overlaid for each oscillation frequency, after they were rescaled by $\nu_\mathrm{Corsaro15}$. The horizontal lines at $(\nu_0 - \nu_\mathrm{Corsaro15}) / \nu_\mathrm{Corsaro15} = 0\,\%$ denote the perfect matching. The panels from left to right depict the result obtained by considering four different methods to extract the frequencies from the counts histogram: \textbf{(a)} using a symmetric frequency range for each local maximum, with an extent on each side of the maximum that is the minimum found between the two distances going from the maximum to the adjacent midpoints from the next and previous maxima (or the edge of the total frequency range) on the right and left side respectively. The final frequencies are computed as the sampling mean from the sampling falling in the selected range of each local maximum, while the corresponding uncertainty is the standard deviation from the same sampling used to compute the mean; \textbf{(b)} using the same symmetric ranges as (a) but this time performing a weighted mean and weighted standard deviation using the nested iteration of each sampling point as a weight; \textbf{(c)} using an asymmetric frequency range, built with the distances between the maximum and the adjacent midpoints from the next and previous maxima (or edge of the total frequency range) on the right and left side respectively. The mean and standard deviation of the selected sampling of the range are used as in (a) to compute the final frequency and uncertainty estimates from each local maximum; \textbf{(d)} same as the case (c) but this time using a weighted mean and standard deviation as done for the case (b).}
\label{fig:2}
\end{figure}

The example shown in Figure~\ref{fig:1} is illustrating the extraction process for a frequency range of length $\Delta\nu$ in the low-luminosity low-mass RGB star KIC~12008916, observed nearly continuously by \kepler\,\,for more than 4 years and using the same PSD as in C15. This star has been adopted by C15 as a reference star to present the results of a peak bagging analysis based on Approach 1. Here I used the same range adopted in Figure 5 of C15 to demonstrate that the new Approach 2 is able to select all the actual oscillation frequency peaks identified by C15 (see Figure~\ref{fig:1} top right panel), with the exception of the only peak that was not deemed significant by C15 according to the peak significance test. For deriving the final frequency estimates and corresponding uncertainties from the counts histogram I have adopted four different definitions, described in Figure~\ref{fig:2}. The different definitions have the purpose of testing the impact of: 1) including a weight in the sample average to improve the frequency accuracy, with a weight given by the number of the nested iteration, meaning that those sampling points that correspond to a higher likelihood value have more weight; 2) adopting a symmetric (or asymmetric) frequency range of the oscillation peak with respect to its highest point (in sampling counts) in order to test how the sampling at low likelihood values can bias the estimate of the oscillation frequency. In Figure~\ref{fig:2} I compare 22 extracted frequencies in the interval $146.5$-$172.0$ $\mu$Hz (length 2$\Delta\nu$) by using the new Approach 2, to those published by C15. The agreement is remarkable for all definitions presented but it is the best when adopting case (b) of Figure~\ref{fig:2} (hence weighting by the nested iteration and adopting the symmetric frequency range), where the new frequencies deviate on average by only 0.01\,\% and have precisions that are on average of about 0.1\,\%. 

It is important to consider that the newly presented Approach 2 has the advantage that can be applied to the complex pattern of the RGB oscillations, as demonstrated in this work, contrary to the original Approach 2 that was limited to main sequence stars only. A direct comparison between the new and old Approach 2 is therefore not possible for this dataset, but one can in general expect to achieve precisions on the frequency estimates that are of the same order of magnitude given that both approaches are based on the sampling of a multi-modal posterior distribution and involve the use of a low number of free parameters.

Lastly, I note that the extraction of the oscillation frequencies using the new Approach 2 takes about 2\,s on a single 2.6 GHz CPU core for the PSD chunk presented in Figure~\ref{fig:1}. This implies that the entire star, containing roughly 70 different oscillation modes, can be processed in about 15\,s. By comparison with Approach 1, where the frequency extraction done by C15 took about 27\,min for the entire star KIC~12008916, we gain an useful factor $\sim100$ in speed.

\section{Discussion}
\label{sec:discussion}
The methodology presented in Section~\ref{sec:fast}, building on a previous result by C14, was able not only to find all the actual oscillation frequency peaks in a range of 25\,$\mu$Hz containing 22 different oscillation modes of KIC~12008916 -- both pressure and mixed dipolar modes and some of which extremely narrow in width and small in height as compared to the background level --  but also to deliver frequencies that are as accurate as 0.01\% for the entire set analyzed, and precise to a level of 0.1\,\%. This implies that these frequencies could be used for modeling purposes, although precisions do not reach the level of those extracted from a peak bagging analysis using Approach 1, which ranges from $10^{-2}$ to $10^{-3}$\,\% for the range of frequencies used in this work, as obtained by C15. This new Approach 2 has two important advantages: 1) it is extremely convenient in terms of computational speed, yielding a factor of about 100 improvement with respect to Approach 1; 2) it does not require any assumption on the location of the frequency peaks, but only a simple average estimate of the oscillation amplitude in a chunk and an estimate of a minimum linewidth for each star, making it an adequate approach also for the complex oscillation patterns found in evolved stars. For the test presented here, I have adopted a FWHM of three times the frequency resolution of the dataset, because this was shown by C15 to be the typical FWHM of a fully unresolved oscillation peak, which is the narrowest observed. Therefore Approach 2 is automated because it does not require any human supervision, with the only requirement of using the global oscillation properties of $\nu_\mathrm{max}$ and $\Delta\nu$ to set up the model and priors, proving that it is perfectly suitable for the analysis of a large sample of targets.

Approach 2 can be easily coupled to Approach 1 in an automated sequence where Approach 2 is used to rapidly define the number of peaks, locate their frequency position and build relatively narrow prior ranges that are then feed in Approach 1 to perform a full peak bagging analysis. As a result, also the peak significance test performed with Approach 1 can be fully automated. This is possible because, thanks to the high accuracy of the $\nu_0$ values extracted with Approach 2, the frequency priors can be set up as the boundaries given by the extracted uncertainties on the frequency centroids from Approach 2, thus eliminating the problem of locating the peaks before performing the actual fit. Future work will aim at investigating the performances of Approach 2, as well as assessing its calibration, in the light of different $\nu_\mathrm{max}$ values, level of background photon noise, actual linewidth of the peaks (e.g. in the presence of blending as for F-type stars), and of different frequency resolutions typical of the new observations from K2 and TESS. 

\section*{Conflict of Interest Statement}
The author declares that the research was conducted in the absence of any commercial or financial relationships that could be construed as a potential conflict of interest.

\section*{Author Contributions}
E.C. has developed the new features implemented in \diamonds, refined Approach 2 from C14, performed its calibration, testing, and the application to the RGB star.

\section*{Funding}
E.C. is funded by the European Union's Horizon 2020 research and innovation program under the Marie Sklodowska-Curie grant agreement No. 664931.

\section*{Acknowledgments}
I thank Dr. Joris De Ridder for helpful discussions.

\section*{Data Availability Statement}
The dataset analyzed for this study can be found in MAST, https://archive.stsci.edu/kepler/.

\bibliographystyle{frontiersinSCNS_ENG_HUMS} 
\bibliography{CORSARO_frontiers}

\begin{thebibliography}{23}
\providecommand{\natexlab}[1]{#1}
\expandafter\ifx\csname urlstyle\endcsname\relax
  \providecommand{\doi}[1]{doi:\discretionary{}{}{}#1}\else
  \providecommand{\doi}{doi:\discretionary{}{}{}\begingroup
  \urlstyle{rm}\Url}\fi
\providecommand{\selectlanguage}[1]{\relax}
\providecommand{\bibAnnoteFile}[1]{%
  \IfFileExists{#1}{\begin{quotation}\noindent\textsc{Key:} #1\\
  \textsc{Annotation:}\ \input{#1}\end{quotation}}{}}
\providecommand{\bibAnnote}[2]{%
  \begin{quotation}\noindent\textsc{Key:} #1\\
  \textsc{Annotation:}\ #2\end{quotation}}

\bibitem[{{Aerts} et~al.(2010){Aerts}, {Christensen-Dalsgaard}, and
  {Kurtz}}]{Aerts10}
{Aerts}, C., {Christensen-Dalsgaard}, J., and {Kurtz}, D.~W. (2010).
\newblock \emph{{Asteroseismology}} (Springer)
\bibAnnoteFile{Aerts10}

\bibitem[{{Baglin} et~al.(2006){Baglin}, {Michel}, {Auvergne}, and {COROT
  Team}}]{Baglin06}
{Baglin}, A., {Michel}, E., {Auvergne}, M., and {COROT Team} (2006).
\newblock {The seismology programme of the CoRoT space mission}.
\newblock In \emph{Proceedings of SOHO 18/GONG 2006/HELAS I, Beyond the
  spherical Sun}. vol. 624 of \emph{ESA Special Publication}, 34
\bibAnnoteFile{Baglin06}

\bibitem[{{Benomar} et~al.(2015){Benomar}, {Takata}, {Shibahashi}, {Ceillier},
  and {Garc{\'{\i}}a}}]{Benomar15}
{Benomar}, O., {Takata}, M., {Shibahashi}, H., {Ceillier}, T., and
  {Garc{\'{\i}}a}, R.~A. (2015).
\newblock {Nearly uniform internal rotation of solar-like main-sequence stars
  revealed by space-based asteroseismology and spectroscopic measurements}.
\newblock \emph{Mon. Not. R. Astron. Soc.} 452, 2654--2674.
\newblock \doi{10.1093/mnras/stv1493}
\bibAnnoteFile{Benomar15}

\bibitem[{{Borucki} et~al.(2010){Borucki}, {Koch}, {Basri}, {Batalha}, {Brown},
  {Caldwell} et~al.}]{Borucki10}
{Borucki}, W.~J., {Koch}, D., {Basri}, G., {Batalha}, N., {Brown}, T.,
  {Caldwell}, D., et~al. (2010).
\newblock {Kepler Planet-Detection Mission: Introduction and First Results}.
\newblock \emph{Science} 327, 977.
\newblock \doi{10.1126/science.1185402}
\bibAnnoteFile{Borucki10}

\bibitem[{{Corsaro}(2018)}]{Corsaro18tutorial}
{Corsaro}, E. (2018).
\newblock {Tutorial: Asteroseismic Data Analysis with DIAMONDS}.
\newblock \emph{Asteroseismology and Exoplanets: Listening to the Stars and
  Searching for New Worlds} 49, 137.
\newblock \doi{10.1007/978-3-319-59315-9_7}
\bibAnnoteFile{Corsaro18tutorial}

\bibitem[{{Corsaro} and {De Ridder}(2014)}]{Corsaro14}
{Corsaro}, E. and {De Ridder}, J. (2014).
\newblock {DIAMONDS: A new Bayesian nested sampling tool. Application to peak
  bagging of solar-like oscillations}.
\newblock \emph{Astron. Astrophys.} 571, A71.
\newblock \doi{10.1051/0004-6361/201424181}
\bibAnnoteFile{Corsaro14}

\bibitem[{{Corsaro} et~al.(2015{\natexlab{a}}){Corsaro}, {De Ridder}, and
  {Garc{\'{\i}}a}}]{Corsaro15cat}
{Corsaro}, E., {De Ridder}, J., and {Garc{\'{\i}}a}, R.~A.
  (2015{\natexlab{a}}).
\newblock {Bayesian peak bagging analysis of 19 low-mass low-luminosity red
  giants observed with Kepler}.
\newblock \emph{Astron. Astrophys.} 579, A83.
\newblock \doi{10.1051/0004-6361/201525895}
\bibAnnoteFile{Corsaro15cat}

\bibitem[{{Corsaro} et~al.(2015{\natexlab{b}}){Corsaro}, {De Ridder}, and
  {Garc{\'{\i}}a}}]{Corsaro15glitch}
{Corsaro}, E., {De Ridder}, J., and {Garc{\'{\i}}a}, R.~A.
  (2015{\natexlab{b}}).
\newblock {High-precision acoustic helium signatures in 18 low-mass
  low-luminosity red giants. Analysis from more than four years of Kepler
  observations}.
\newblock \emph{Astron. Astrophys.} 578, A76.
\newblock \doi{10.1051/0004-6361/201525922}
\bibAnnoteFile{Corsaro15glitch}

\bibitem[{{Corsaro} et~al.(2018){Corsaro}, {De Ridder}, and
  {Garc{\'{\i}}a}}]{Corsaro18corr}
{Corsaro}, E., {De Ridder}, J., and {Garc{\'{\i}}a}, R.~A. (2018).
\newblock {Bayesian peak bagging analysis of 19 low-mass low-luminosity red
  giants observed with Kepler (Corrigendum)}.
\newblock \emph{Astron. Astrophys.} 612, C2.
\newblock \doi{10.1051/0004-6361/201525895e}
\bibAnnoteFile{Corsaro18corr}

\bibitem[{{Corsaro} et~al.(2017{\natexlab{a}}){Corsaro}, {Lee},
  {Garc{\'{\i}}a}, {Hennebelle}, {Mathur}, {Beck} et~al.}]{Corsaro17Nature}
{Corsaro}, E., {Lee}, Y.-N., {Garc{\'{\i}}a}, R.~A., {Hennebelle}, P.,
  {Mathur}, S., {Beck}, P.~G., et~al. (2017{\natexlab{a}}).
\newblock {Spin alignment of stars in old open clusters}.
\newblock \emph{Nat. Astron.} 1, 0064.
\newblock \doi{10.1038/s41550-017-0064}
\bibAnnoteFile{Corsaro17Nature}

\bibitem[{{Corsaro} et~al.(2017{\natexlab{b}}){Corsaro}, {Mathur},
  {Garc{\'{\i}}a}, {Gaulme}, {Pinsonneault}, {Stassun}
  et~al.}]{Corsaro17metallicity}
{Corsaro}, E., {Mathur}, S., {Garc{\'{\i}}a}, R.~A., {Gaulme}, P.,
  {Pinsonneault}, M., {Stassun}, K., et~al. (2017{\natexlab{b}}).
\newblock {Metallicity effect on stellar granulation detected from oscillating
  red giants in open clusters}.
\newblock \emph{Astron. Astrophys.} 605, A3.
\newblock \doi{10.1051/0004-6361/201731094}
\bibAnnoteFile{Corsaro17metallicity}

\bibitem[{{Davies} et~al.(2016){Davies}, {Silva Aguirre}, {Bedding},
  {Handberg}, {Lund}, {Chaplin} et~al.}]{Davies16}
{Davies}, G.~R., {Silva Aguirre}, V., {Bedding}, T.~R., {Handberg}, R., {Lund},
  M.~N., {Chaplin}, W.~J., et~al. (2016).
\newblock {Oscillation frequencies for 35 Kepler solar-type planet-hosting
  stars using Bayesian techniques and machine learning}.
\newblock \emph{Mon. Not. R. Astron. Soc.} 456, 2183--2195.
\newblock \doi{10.1093/mnras/stv2593}
\bibAnnoteFile{Davies16}

\bibitem[{{Garc{\'{\i}}a Saravia Ortiz de Montellano}
  et~al.(2018){Garc{\'{\i}}a Saravia Ortiz de Montellano}, {Hekker}, and
  {Theme{\ss}l}}]{Garcia18pb}
{Garc{\'{\i}}a Saravia Ortiz de Montellano}, A., {Hekker}, S., and
  {Theme{\ss}l}, N. (2018).
\newblock {Automated asteroseismic peak detections}.
\newblock \emph{Mon. Not. R. Astron. Soc.} 476, 1470--1496.
\newblock \doi{10.1093/mnras/sty253}
\bibAnnoteFile{Garcia18pb}

\bibitem[{{Handberg} et~al.(2017){Handberg}, {Brogaard}, {Miglio}, {Bossini},
  {Elsworth}, {Slumstrup} et~al.}]{Handberg17}
{Handberg}, R., {Brogaard}, K., {Miglio}, A., {Bossini}, D., {Elsworth}, Y.,
  {Slumstrup}, D., et~al. (2017).
\newblock {NGC 6819: testing the asteroseismic mass scale, mass loss and
  evidence for products of non-standard evolution}.
\newblock \emph{Mon. Not. R. Astron. Soc.} 472, 979--997.
\newblock \doi{10.1093/mnras/stx1929}
\bibAnnoteFile{Handberg17}

\bibitem[{{Howell} et~al.(2014){Howell}, {Sobeck}, {Haas}, {Still}, {Barclay},
  {Mullally} et~al.}]{Howell14K2}
{Howell}, S.~B., {Sobeck}, C., {Haas}, M., {Still}, M., {Barclay}, T.,
  {Mullally}, F., et~al. (2014).
\newblock {The K2 Mission: Characterization and Early Results}.
\newblock \emph{Pub. Astron. Soc. Pac.} 126, 398--408.
\newblock \doi{10.1086/676406}
\bibAnnoteFile{Howell14K2}

\bibitem[{{Keeton}(2011)}]{Keeton11}
{Keeton}, C.~R. (2011).
\newblock {On statistical uncertainty in nested sampling}.
\newblock \emph{Mon. Not. R. Astron. Soc.} 414, 1418--1426.
\newblock \doi{10.1111/j.1365-2966.2011.18474.x}
\bibAnnoteFile{Keeton11}

\bibitem[{{Koch} et~al.(2010){Koch}, {Borucki}, {Basri}, {Batalha}, {Brown},
  {Caldwell} et~al.}]{Koch10}
{Koch}, D.~G., {Borucki}, W.~J., {Basri}, G., {Batalha}, N.~M., {Brown}, T.~M.,
  {Caldwell}, D., et~al. (2010).
\newblock {Kepler Mission Design, Realized Photometric Performance, and Early
  Science}.
\newblock \emph{Astrophys. J. Lett.} 713, L79-L86.
\newblock \doi{10.1088/2041-8205/713/2/L79}
\bibAnnoteFile{Koch10}

\bibitem[{{Lund} et~al.(2017){Lund}, {Silva Aguirre}, {Davies}, {Chaplin},
  {Christensen-Dalsgaard}, {Houdek} et~al.}]{Lund17LEGACY}
{Lund}, M.~N., {Silva Aguirre}, V., {Davies}, G.~R., {Chaplin}, W.~J.,
  {Christensen-Dalsgaard}, J., {Houdek}, G., et~al. (2017).
\newblock {Standing on the Shoulders of Dwarfs: the Kepler Asteroseismic LEGACY
  Sample. I. Oscillation Mode Parameters}.
\newblock \emph{Astrophys. J.} 835, 172.
\newblock \doi{10.3847/1538-4357/835/2/172}
\bibAnnoteFile{Lund17LEGACY}

\bibitem[{{Rauer} et~al.(2014){Rauer}, {Catala}, {Aerts}, {Appourchaux},
  {Benz}, {Brandeker} et~al.}]{Rauer14PLATO}
{Rauer}, H., {Catala}, C., {Aerts}, C., {Appourchaux}, T., {Benz}, W.,
  {Brandeker}, A., et~al. (2014).
\newblock {The PLATO 2.0 mission}.
\newblock \emph{Experimental Astronomy} 38, 249--330.
\newblock \doi{10.1007/s10686-014-9383-4}
\bibAnnoteFile{Rauer14PLATO}

\bibitem[{{Ricker} et~al.(2014){Ricker}, {Winn}, {Vanderspek}, {Latham},
  {Bakos}, {Bean} et~al.}]{Ricker14TESS}
{Ricker}, G.~R., {Winn}, J.~N., {Vanderspek}, R., {Latham}, D.~W., {Bakos},
  G.~{\'A}., {Bean}, J.~L., et~al. (2014).
\newblock {Transiting Exoplanet Survey Satellite (TESS)}.
\newblock In \emph{Society of Photo-Optical Instrumentation Engineers (SPIE)
  Conference Series}. vol. 9143 of \emph{Society of Photo-Optical
  Instrumentation Engineers (SPIE) Conference Series}, 20.
\newblock \doi{10.1117/12.2063489}
\bibAnnoteFile{Ricker14TESS}

\bibitem[{Sivia and Skilling(2006)}]{Sivia06}
Sivia, D. and Skilling, J. (2006).
\newblock \emph{Data Analysis: A Bayesian Tutorial}.
\newblock Oxford science publications (OUP Oxford)
\bibAnnoteFile{Sivia06}

\bibitem[{Skilling(2004)}]{Skilling04}
Skilling, J. (2004).
\newblock Nested sampling.
\newblock \emph{AIP Conference Proceedings} 735, 395 (SK04)--405.
\newblock \doi{http://dx.doi.org/10.1063/1.1835238}
\bibAnnoteFile{Skilling04}

\bibitem[{{Vrard} et~al.(2018){Vrard}, {Kallinger}, {Mosser}, {Barban},
  {Baudin}, {Belkacem} et~al.}]{Vrard18pb}
{Vrard}, M., {Kallinger}, T., {Mosser}, B., {Barban}, C., {Baudin}, F.,
  {Belkacem}, K., et~al. (2018).
\newblock {Amplitude and lifetime of radial modes in red giant star spectra
  observed by Kepler}.
\newblock \emph{Astron. Astrophys.} 616, A94.
\newblock \doi{10.1051/0004-6361/201732477}
\bibAnnoteFile{Vrard18pb}

\end{thebibliography}

\end{document}